\documentclass[fleqn,10pt]{wlscirep}

\title{The Beneficial Role of Mobility for the Emergence of Innovation}

\author[1]{Giuliano Armano}
\author[2,*]{Marco Alberto Javarone}
\affil[1]{Department of Electronics and Computer Engineering, University of Cagliari, Cagliari 09123 - Italy}
\affil[2]{Department of Mathematics and Computer Science, University of Cagliari, Cagliari 09123 - Italy}

\affil[*]{marcojavarone@gmail.com}

\begin{abstract}
Innovation is a key ingredient for the evolution of several systems, including social and biological ones. Focused investigations and lateral thinking may lead to innovation, as well as serendipity and other random discovery processes. Some individuals are talented at proposing innovation (say innovators), while others at deeply exploring proposed novelties, at getting further insights on a theory, or at developing products, services, and so on (say developers). This separation in terms of innovators and developers raises an issue of paramount importance: under which conditions a system is able to maintain innovators? According to a simple model, this work investigates the evolutionary dynamics that characterize the emergence of innovation. In particular, we consider a population of innovators and developers, in which agents form small groups whose composition is crucial for their payoff. The latter depends on the heterogeneity of the formed groups, on the amount of innovators they include, and on an award-factor that represents the policy of the system for promoting innovation. Under the hypothesis that a ''mobility'' effect may support the emergence of innovation, we compare the equilibria reached by our population in different cases. Results confirm the beneficial role of ''mobility'', and the emergence of further interesting phenomena. 
\end{abstract}
\begin{document}

\flushbottom
\maketitle
\thispagestyle{empty}


\section*{Introduction}

Innovation~\cite{johnson01} appears to be an ubiquitary concept, which applies to a variety of contexts, including economy, physics, sociology, ethology, biology, and linguistics~\cite{loreto01,galla01,amaral01,pietronero01,loreto02,thurner01,tether01,valente01,rogers01,reader01,brien01,loreto03,javarone01,javarone02}.
A typical setting able to support innovation requires a component, e.g., a research group, whose specific goal is to produce a breakthrough --–which in turn is a precondition to find out new technologies, services, and even forms of art~\cite{tschmuck01}. As such, the capability of producing innovation becomes also an indicator of the wellness of a society~\cite{ziman01,sole01,zabell01}.
On the other hand, in its pure form, innovation is in fact an unexpected outcome, most likely due to random guessing, lateral thinking or serendipity. The most prominent examples of this kind of mechanism can be found in science, where innovation is fundamental for promoting groundbreaking intuitions~\cite{sinatra01}. In this context innovation is motivated by the goal of dealing with unsolved problems and sometimes carries out, as side-effect, the emergence of new research fields. A relevant and recent example is constituted by the modern and vibrant field of complex networks~\cite{barabasi01,newman01,caldarelli01,estrada01}, which is deeply affecting several scientific sectors (just to cite few among many: social networks~\cite{leskovec01}, epidemiology~\cite{vespignani01}, genomics~\cite{sole02}, neuroscience~\cite{sporn01,marinazzo01}, and financial systems~\cite{battiston01}).
Regardless from the underlying context, the ability of identifying objective measures for innovation is a primary issue. A relevant discipline for this purpose is scientometrics, which is concerned with studying, measuring and analysing science, technology and innovation. Narrowing to scientific publications, scientometrics is the reference discipline for measuring their impact on the corresponding research field. A further concern with scientometrics is studying the emergence and the evolution of scientific collaborations~\cite{perc04,sun01,goldman01,latora01}.
It is worth pointing out that innovation is not specifically tied to human activities, as ''by design'' all biological systems are able to support it. In biology, innovation pertains to it from both structural and behavioral perspectives. In the former case, one may observe (or infer) the emergence of new gene sequences in living organisms able to improve their fitness. In the latter case, specific studies concerning animal behavior pointed out that also animals are able to come up with innovative solutions~\cite{reader01}.
Remarkably, also evolutionary computation (in particular, genetic algorithms~\cite{goldberg01}), strongly emphasizes the role of innovation versus development (in the jargon used in that research community: exploration versus exploitation~\cite{marjan01}) while evolving bit strings according to an oversimplification of the general principles that hold for the evolution of biological systems.

In the light of these observations, we deem that the emergence of innovation can be viewed as an evolutionary process where several actors are involved (see~\cite{valverde01}). In particular, some actors propose new ideas, while others develop these ideas turning them into practical technologies, services, and so on. As a matter of fact, the actual characteristics of an individual typically lay in the middle between innovators and developers. However, to better investigate the coexistence between these characters, we assume that an individual can be either innovator or developer. Besides, a similar view has also been proposed in the field of mathematics by Dyson~\cite{dyson01}, who divided mathematicians in two groups: birds and frogs.
According to his picture, the former fly high in the air and survey broad vistas of mathematics out to the far horizons, and become aware about the connections between different fields, whereas the latter --from their position-- are able to appreciate with more detail the flowers that grow nearby, i.e. they have a more granular and fine views of mathematical concepts and theories.
This work is aimed at investigating the tight relationship that holds between innovators and developers, studying the underlying process using the Evolutionary Game Theory framework~\cite{perc01,perc02,perc03,nowak01,nowak02,moreno01,moreno02,santos01,santos02,szabo01,szabo02,javarone03,moreno03,antonioni01} (EGT hereinafter).
To this end a specific game, named ''Innovation Game'', has been set up for shading light on the equilibria reached by a population composed of innovators and developers. Before to proceed, showing results of our investigations, we deem interesting to mention some previous attempts in describing the emergence of innovation, even if from a different point of view, that were based on EGT, i.e.~\cite{szolnoki10,amaral02,szabo03}.

\section*{Results}

\begin{figure}[h]
\centering
\includegraphics[width=0.55\textwidth]{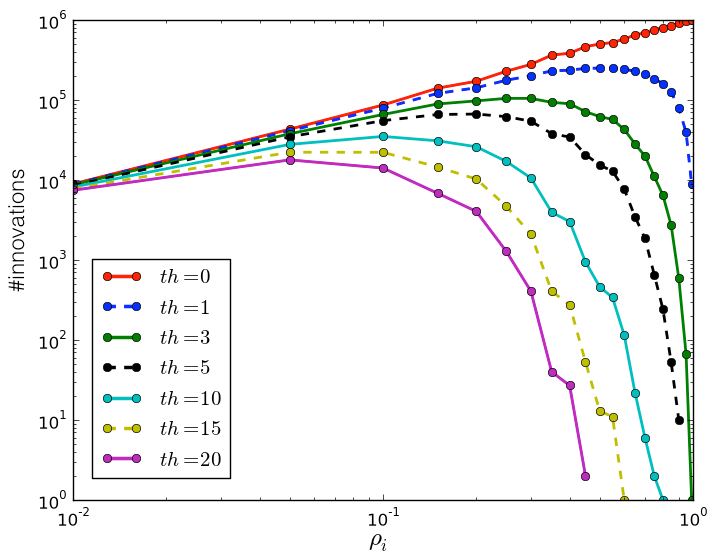}
\caption{\small Number of innovations as function of innovators' density (i.e. $\rho_i$) in a population with $N = 1000$ agents. Each curve illustrates results achieved using a different threshold $Th$. This threshold is used to quantify the amount of innovation of a proposal (e.g., the number of citations in a scientific context or the number of implementations of a new technology). The main role of $Th$ is to link the emergence of novelties with the existence of a community of developers able to put them into practice. Results have been averaged over different simulation runs. \label{fig:innovators_function}}
\end{figure}

The first issue that has been tackled was about the underlying context within which interactions between innovators and developers are supposed to occur. In fact, despite the availability of many tools that have been devised in support of collaborative work, apparently the most effective collaborations among humans still occurs on a local basis. As a consequence, our population splits into small groups of fixed size, though preserving the possibility of rendering pseudo-random groups. In so doing, we admit the possibility of ensuring mobility among agents, depending on the adopted grouping strategy. In either case, the density of innovators in a group ($\rho_i$, hereinafter) constitutes a parameter of the system.
We then concentrated on how to model the presence of innovators and developers in a group. Let us briefly summarize the concerns about innovators and developers. As for innovators, although their presence is mandatory to get new insights, we had to consider the fact that they often represent a risk ---for they may be not successful at all over long periods of time. To account for both aspects, we introduced an \textit{award factor}, aimed at accounting for the benefit for including innovators in a group, together with a penalty, aimed at accounting for the cost of unsuccessful insights (or no insights at all) over time. 
As for developers, their modeling did not require any specific care, as they typically tend to be effective from the very beginning of any activity they are involved in (e.g., a research project), and tend to establish tight relationships with their neighbors and with the hosting structure as well.
Then, we had to model the way innovative thinking can propagate over the given population. To better understand the underlying issues, let us consider the relevant and well-known case concerning scientific publications. In this case, the whole process starts with publishing the results concerning a novel insight, or an improvement over an existing idea, in a scientific journal. Depending on the degree of “penetrance” of the published paper, it may undergo citations and the underlying idea or technology may be further improved by other people. Hence, at least in principle, one can evaluate publications in accordance with the amount of citations. It is worth to clarify that receiving a citation (i.e. a mention and/or further attention) does not imply that an idea is more important than another. It just means that a community, according to its guidelines and rules, decides to follow and investigate specific ideas rather than others (let us recall, for instance, that both Einstein’s General Relativity and the Higgs’s boson required a lot of time before being accepted and recognized as real breakthroughs). Notwithstanding the peculiarities related to the time required for a novel insight to be accepted and disseminated within the scientific community, in either case the number of novelties appears to be tightly related with the amount of innovators.
Fig.~\ref{fig:innovators_function} shows the number of novelties as function of the density of innovators in a population composed also by developers. A threshold (say $Th$) has been adopted to deal with the problem of discriminating between proposals perceived as interesting (beyond their actual relevance) and proposals that are not. For instance, in the scientific community, this threshold may represent the minimum number of citations required to claim that a manuscript has been able to improve the state of the art.
Going back to our attempt to model the innovation game, it should be now clearer why the density of innovators $\rho_i$ becomes a crucial parameter of the model. In this scenario, one may wonder whether increasing the mobility among agents increases the density of innovators (while keeping fixed the amount of available resources) in a system where each agent can change its behavior according to a gained payoff, i.e. being driven by a rational mindset. 
This question is also motivated by qualitative observations of the real world. For instance, many researchers like to spend part of their time visiting external labs, and in general workers often change companies also for improving their experience and skills. In addition, also biology suggests that mobility can be helpful for new solutions, as marriages between individuals without any degree of kinship reduce the probability to transmit diseases to their offspring. 
Eventually, as reported below, results confirmed our hypothesis, also suggesting that the emergence of innovation depends on the amount of resources assigned to it.

Summarizing, the proposed innovation game occurs within a population of agents organized in groups of size $G$. 
Here, the payoff of each agent depends on the following factors: the heterogeneity of the formed groups, the number of innovators $I$ in these groups, and an award factor $aw$. The latter is a numerical parameter that represents the efforts made by a system for promoting innovation.
Hence, the payoff can be defined as follows:
\begin{equation}\label{eq:payoff}
\pi = aw \cdot (1 - \frac{M}{2}) - \frac{I}{G}
\end{equation}
\noindent where the $M$ magnetization factor~\cite{huang01} has been introduced for measuring the group heterogeneity.
With $D$ number of developers in a group, we can write:
\begin{equation}\label{eq:magnetization}
M = \frac{|I - D|}{G}
\end{equation}
Assigning a spin equal to $+1$ to innovators and a spin equal to $-1$ to developers, the value of $M$ falls in the range $[0,1]$.
The population evolves according to the following dynamics: at each time step, one agent (say $x$) is randomly selected, together with one of its neighbors (say $y$). Moreover, $x$ and $y$ receive a payoff (say $\pi_x$ and $\pi_y$, respectively) according to the definition given in eq.~\eqref{eq:payoff}.
Then, the agent $y$ imitates the strategy of the agent $x$ (see also~\cite{miguel02}) with a probability that depends on the difference between their payoffs, the greater $\pi_x - \pi_y$, the greater the probability that $y$ imitates $x$. Notably, this probability is computed by the following Fermi-like function:
\begin{equation}\label{eq:prob_transition}
W(s^y \leftarrow s^x) = \left(1 + \exp\left[\frac{\pi^y - \pi^x}{K}\right]\right)^{-1}
\end{equation}
\noindent where $s^x$ and $s^y$ denote the strategy adopted by agent $x$ and $y$, respectively, whereas $K$ represents noise (set to $0.5$, see~\cite{perc01} for further details).
Eq.~\eqref{eq:payoff} is actually embedding two different parts, i.e. $aw \cdot (1 - M/2)$ and $-I/G$. The former allows to promote heterogeneous groups, as the magnetization $M$ goes to $0$ when the amount of innovators approaches that of developers in the same group. The latter represents the additional fee (i.e. the cost) due to the presence of innovators.
From a statistical physics point of view~\cite{huang01,barra01}, the dynamics of the population can be assimilated to that of a spin system (e.g. see~\cite{barra02}), so that at high temperature one expects a disordered (paramagnetic) phase, while at low temperatures (i.e. lower than the critical one $T_c$) one expects an ordered (ferromagnetic) phase.

\begin{figure*}[h]
\centering
\includegraphics[width=1.00\textwidth]{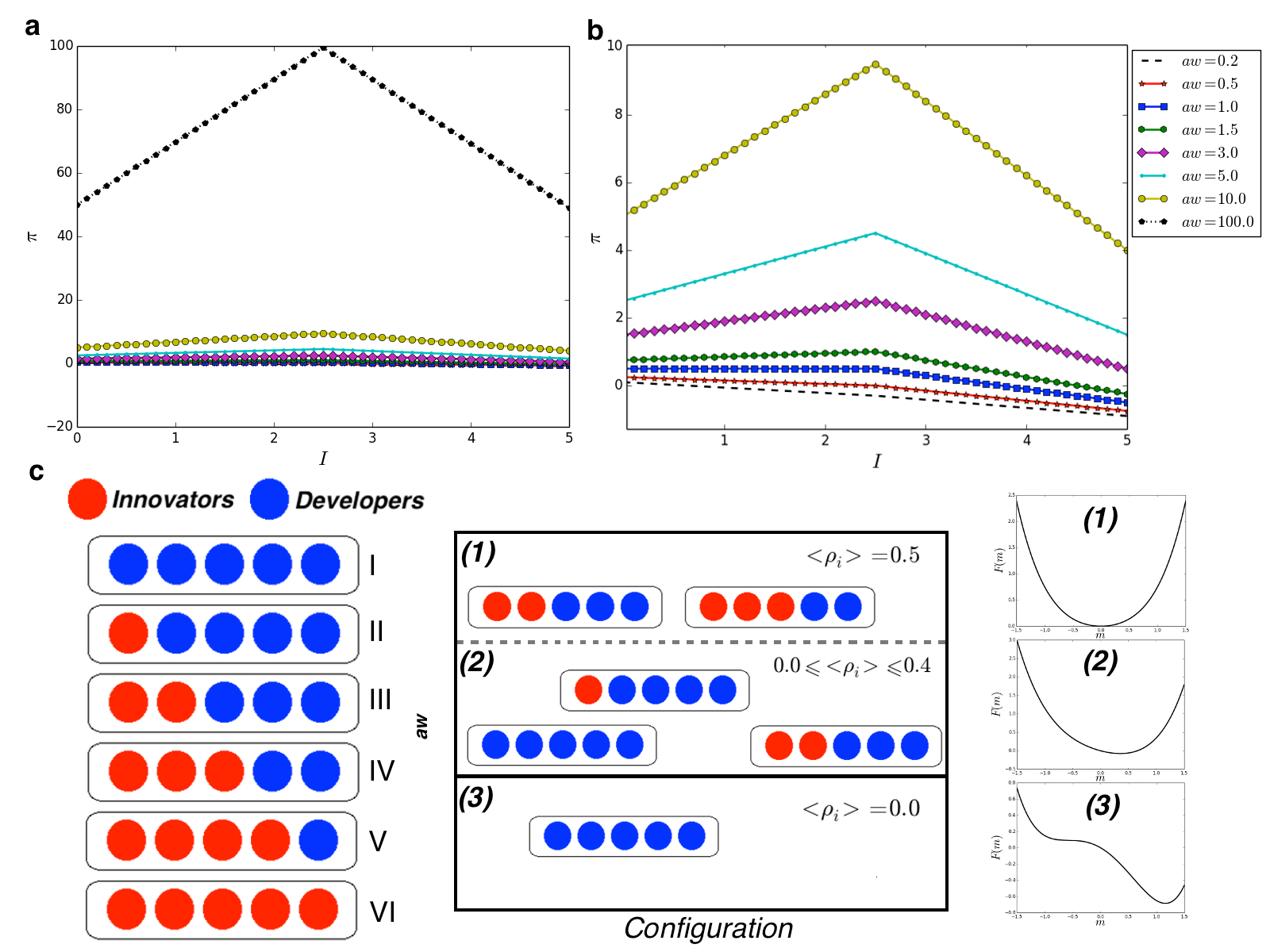}
\caption{\small \textbf{a}) Payoff $\pi$ in function of the number of innovators $I$ in a group, for different values of the awarding factor $aw$ (see legend). \textbf{b}) Enlarged view of the plot shown in \textbf{a}), for values of $aw$ in the range $[0.2, 10.0]$. \textbf{c}) Pictorial representation of Innovators (Red) and Developers (Blue), distributed in groups with different combinations. Each combination, identified in terms of the average density of innovators $<\rho_i>$, can be found in a particular equilibrium. The latter depends on the value of $aw$.  \label{fig:basic}}
\end{figure*}

\begin{figure}[h]
\centering
\includegraphics[width=0.55\textwidth]{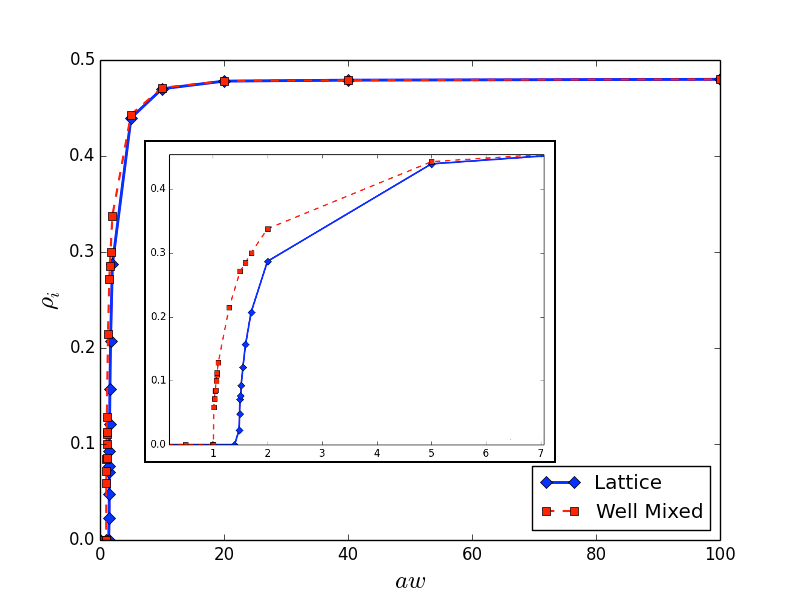}
\caption{\small Density of innovators varying the \textit{award factor} in both configurations, i.e. lattice and well mixed. \label{fig:density_innovators}}
\end{figure}

Notably, in our population, the paramagnetic phase corresponds to the presence of both kinds of agents, whereas the ferromagnetic one corresponds to the presence of only one species.
Thus, at low temperatures, considering a payoff due only to the left-hand part of eq.~\eqref{eq:payoff} and a mixed population with a density of innovators equal to $0.5$ at $t=0$, the expected equilibrium corresponds to an ordered phase in which both species can prevail with equal probability. Conversely, due to the presence of the right-hand part of eq.~\eqref{eq:payoff}, the expected equilibrium at low temperatures corresponds to a population composed of developers only (i.e., such that $I = 0$). Then, at high temperatures, the expected equilibrium is a phase where both species coexist.
In our view, the \textit{award factor} can be mapped to the temperature of a spin system. In doing so, for different values of $aw$, we can draw the resulting agent combinations in a group with $G = 5$ individuals  ---see fig.~\ref{fig:basic}.
Remarkably, the right part of eq.~\eqref{eq:payoff} plays the role of ''field generator''. Indeed, for low values of $aw$, there is only one possible ordered equilibrium, as observed in a spin system at low temperatures in presence of an external field.
A more complete view is shown in plot \textbf{c} of fig.~\ref{fig:basic}, which reports a scheme $aw$ versus $configuration$, where three different phases, denoted as \textbf{(1)}, \textbf{(2)}, and \textbf{(3)}, are identified. Clearly, here it is worth to clarify that both \textbf{(1)} and \textbf{(2)} correspond to a paramagnetic phase (as shown on the right-hand side of the schema, which illustrates a pictorial representation of the free energy~\cite{huang01}). 
In particular, phase \textbf{(1)} can be achieved only for high values of $aw$, when the effect of the 'field generator' becomes too small for affecting the system, i.e. the imitation process. Notably, as shown in plots~\textbf{a} and \textbf{b} of fig.~\ref{fig:basic}, increasing $aw$ the payoff becomes a symmetric function around the value $I = 2.5$. This value represents the case with an equal amount of innovators and developers.   
Then, according to a preliminary overview driven by statistical physics, we aim to characterize the phase transition occurring in our population on varying its temperature, i.e. the \textit{award factor}. Now, results can be related to the scenario represented in fig.~\ref{fig:density_innovators}, which points out the expected number of novelties (e.g., original ideas) over time according to the given amount of resources.

\begin{figure}[h]
\centering
\includegraphics[width=0.55\textwidth]{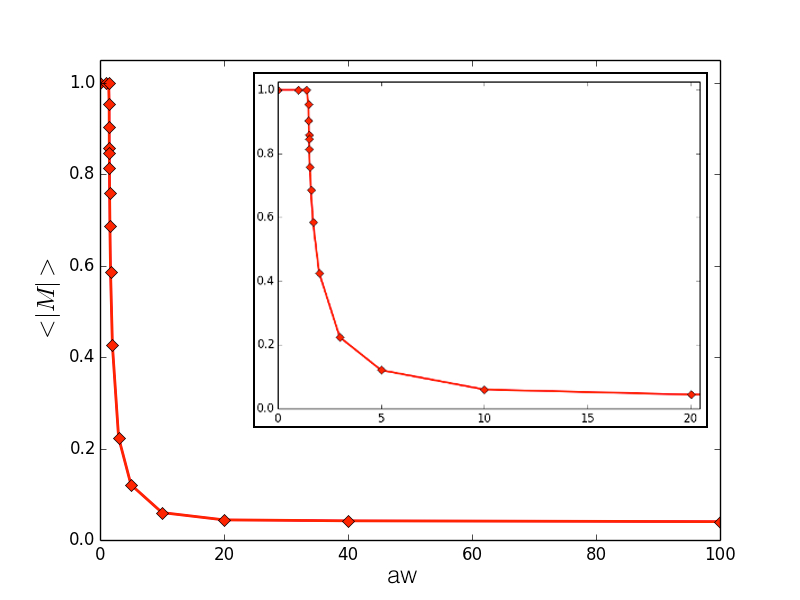}
\caption{\small Average magnetization $|M|$ in function of $aw$. The inset shows a focus of $|M|$ in the small range close to the critical value, i.e.~$aw_c$. \label{fig:magnetization}}
\end{figure}

To investigate the dynamics of the proposed model and its equilibria, we performed numerical simulations.
To this end, we considered two different configurations: a well-mixed population and a population arranged on a square lattice with periodic boundary conditions. Considering a number of agents $N = 10^4$, the side of the square lattice is $L = 100$. In addition, setting $G = 5$, each agent in the lattice belongs to $5$ different groups, therefore also for the well-mixed case we consider this scenario.
Now, it is important to emphasize that comparing the dynamics of the model in the two described configurations has two main motivations. First, we can evaluate if the ''network-reciprocity'' effect~\cite{nowak03} supports innovation, as it does for the cooperation in dilemma games (e.g.,~\cite{perc02}). Second, the well-mixed case allows to represent a sort of ''mobility'' effect, so that we can evaluate its influence in the dynamics of the population, which can be helpful to shade light on the hypothesis that mobility is able to support innovation.
Figure~\ref{fig:density_innovators} shows the density of innovators $\rho_i$, at equilibrium (or after $10^8$ time steps), in function of $aw$.
As expected, for high values of $aw$, both configurations show that the density of innovators slowly tends to $0.5$. Instead, for low values of $aw$ we find a critical threshold $aw_c$. Remarkably, in the well-mixed case, $aw_c$ is smaller than in the lattice topology, indicating that with poor resources, innovators survive only when some kind of mobility is introduced in the system.
\begin{figure}[h]
\centering
\includegraphics[width=1.0\textwidth]{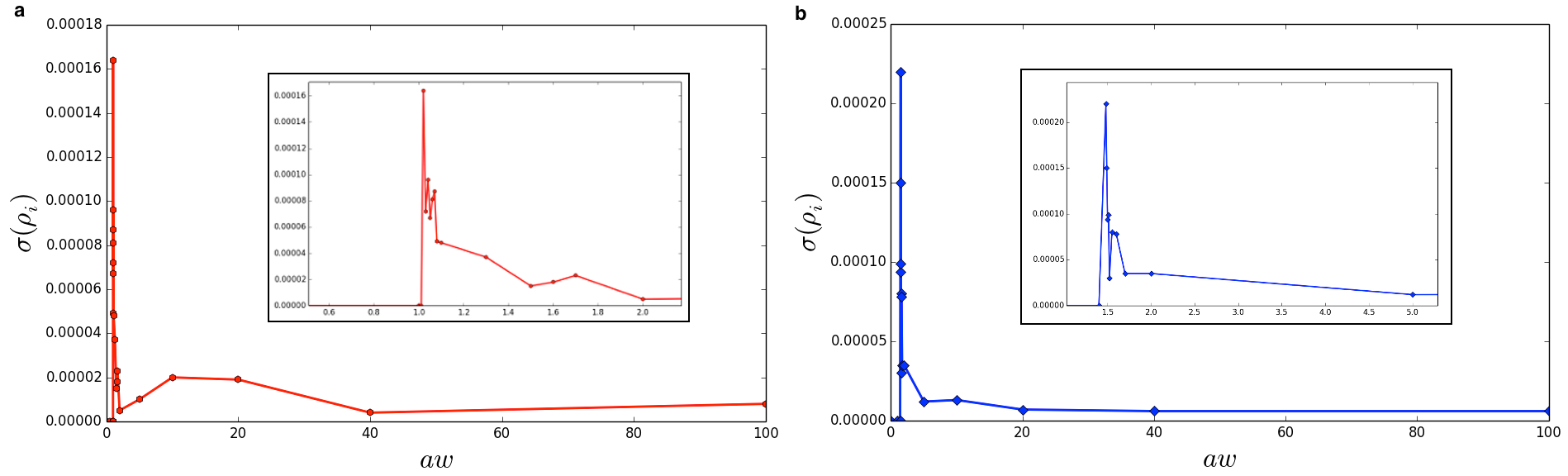}
\caption{\small Variance of $\rho_i$, i.e. $\sigma (\rho_i)$ in function of $aw$, for both configurations: \textbf{a}) well mixed and \textbf{b}) regular lattice. \label{fig:variance_innovators}}
\end{figure}
Then, as reported in fig.~\ref{fig:variance_innovators}, we studied the variance of $\rho_i$, confirming the relevance of the critical values observed in fig.~\ref{fig:density_innovators}.
With the aim to characterize, at least qualitatively, the nature of the transition occurring in our population tuning the value of $aw$, we studied the system magnetization ---see fig.~\ref{fig:magnetization}. Notably, the latter is an order parameter~\cite{huang01}, whose relation with the temperature $T$ (represented by $aw$ in the proposed model), is well known in statistical physics --although often difficult to quantify.
A glance to the system magnetization suggests that the observed phase transition can be classified as of first-order. In particular, with $aw$ approaching $aw_c$, innovators are able to survive and quickly reach a density similar to that of developers.
As final note, let us point out that the group size $G$ is a further parameter that might deserve attention while defining the payoff (i.e.~\ref{eq:payoff}). Indeed, as reported in previous works~\cite{moreno03,perc15,wu01}, the group size may strongly affect the dynamics of the population when playing evolutionary games.
Here, the probability to form heterogeneous groups increases as the size of $G$ increases. As a consequence, the critical value of $aw$ should reduce while increasing $G$. To investigate this issue, we performed some preliminary experiments in the regular lattice. The corresponding results show that with $G = 9$ the critical value for $aw$ reduces from $1.48$ (achieved with $G = 5$) to $1.38$. Notably, these outcomes are in full accordance with the previous aforementioned investigations, aimed at understanding the influence of the groups size in the dynamics of a population.

\section*{Discussion}

This work studies the emergence of innovation under the framework of evolutionary game theory. In particular, by means of a model, inspired to the well-known Dyson’s~\cite{dyson01} classification of mathematicians (see also~\cite{sinatra02}), we analyze the dynamics of a population in terms of innovators and developers. In the proposed model innovators are expected to generate benefits, although they represent also a risk –--as their work can sometimes be unsuccessful.
In order to investigate all related issues, we defined a game where agents form small groups (see also~\cite{szolnoki02,galam02,galam03,bessi01}). Their payoff depends on the heterogeneity of these groups (see~\cite{javarone04}), on the amount of innovators, and on an award factor. Notably, heterogeneity supports the emergence of groups composed of both kinds of agents. The amount of innovators is controlled by an additional parameter, which basically accounts for the fact that they may be unsuccessful for long periods of time. Eventually, the award factor represents the policy of a system in favor of innovation. It is worth recalling that the award factor plays a role similar to that of the synergy factor used in the Public Goods Game for promoting cooperation~\cite{perc03}.
After providing a brief overview of the proposed model inspired by statistical physics, mainly based on the structure of the payoff (eq.~\eqref{eq:payoff}), we performed many numerical simulations. In particular, two main configurations have been investigated: square lattice, with periodic boundary conditions, and well-mixed populations. Notably, the latter allows to represent a mobility effect, which is very important in several real contexts.
Simulations showed that the ''network reciprocity'' effect~\cite{nowak03}, useful for promoting cooperation in many dilemma games having a Nash equilibrium of defection, here reduces the amount of innovators.
It is worth to highlight that, in the proposed model, agents can interact only when they belong to the same group, no matter whether they are directly connected or not in a lattice.
On the other hand, topological interactions allow only to perform imitation processes, i.e. in a structured population one agent can imitate only its nearest-neighbors.
Eventually, we found that the mobility effect plays a beneficial role for supporting innovation, in particular when poor resources (e.g., financial ones) are reserved for innovation. 
Going forward to real systems, we deem that our results may be a starting point to provide general interpretations of well known phenomena. In fact, the density of innovators could even be interpreted as the fraction of time allowed to people, working in a company or institution, to devise new projects and/or ideas. Notable examples in support of this interpretation are some online services provided by Google (e.g., Gmail~\cite{wiki}), which have been devised and designed by collaborators that were allowed to spend a fraction of their working time on new and independent projects.
Finally, we studied the order-disorder phase transition occurring in our model, computing the critical thresholds of the award factor (i.e. $aw_c$). The latter allows to link the model to a simpler scenario that investigates whether a trade-off between innovators and developers in a society can be found (as shown in fig.~\ref{fig:innovators_function}).
It is worth highlighting that, from a game theory perspective, our model is not a dilemma game, like for instance the Prisoner's Dilemma~\cite{moreno02,perc02}. In other words, here agents do not take decisions choosing between their own benefit and that of their community.
Before concluding, let us to spend few words about a possible experimental validation of our model. It is well known that, in the era of Big Data~\cite{madden01,perra01}, the access to real data may allow to verify the usefulness of theoretical models. Remarkably, while the latter allow to speculate about the nature of a phenomenon, often providing important insights, direct investigations, based on real data, typically allow to confirm (or confute) theories, and often open the way to new developments.
In this work, we followed only a theoretical approach, based on qualitative observations of real scenarios. However, notwithstanding the fact that the model proposed in this work has not yet been validated with real-world data, let us briefly describe the kinds of dataset that might be suitable to this extent. In our view, a main experimental scenario would be the world of academia. In principle, every scientific paper that describes an original research represents an innovation. Hence, a dataset where research grants, and other forms of support, related to a group, to a department or to an university, could be very helpful to investigate the correlation between funding, publication of scientific articles and innovation. In this context, a more specific dataset, e.g., describing the amount of funding devised for exchange programs (as visiting professorship and/or studentship), would allow to evaluate if the proposed model can fit real-word data on mobility and its capability to promote innovation.
Considering industry, real data related to the investments of companies on innovation, as well as in supporting employers to attend workshops and conferences, could be useful. Obviously, in this case, real exchange programs (as those that hold for academia) often cannot be implemented for a number of reasons. However, promoting the attendance to events like workshops might be considered, to some extent, as a kind of mobility. Here, the task would be to identify relations between the amount of financial resources reserved for mobility (as above described), by a company and the level of innovation of its products or services. The amount of time granted to collaborators for developing independent projects could be very useful as well.
As for future work, we aim to analyze the proposed model by arranging agents on more complex topologies, e.g., scale-free networks.

\section*{Methods}

Numerical simulations have been performed on a square lattice with periodic boundary conditions and in a non-structured population. So, all agents have a degree equal to $4$, i.e. they have four nearest-neighbors.
In both configurations, agents form $5$ groups of size $G = 5$, and the population evolves according to the following dynamics:
\begin{enumerate}
\item Define a population with $N$ agents, with the same amount of Innovators and Developers (i.e. $\rho_i(0)$ = 0.5);
\item Randomly select an agent, say $x$, and one of its neighbors, say $y$;
\item According to the groups they belong, $x$ and $y$ receive a payoff, i.e. $\pi_x$ and $\pi_y$, respectively;
\item Agent $y$ imitates the strategy of $x$ according to the probability defined in eq.~\eqref{eq:prob_transition};
\item Repeat from $(2)$, until the population reaches an ordered phase or the maximum number of time steps $T$ has been reached.
\end{enumerate}
The ordered phase mentioned in step $(5)$ indicates configurations in which agents follow the same strategy/behavior. The maximum number of time steps has been set to $T = 10^8$.

\section*{Acknowledgements}

The authors wish to thank Luciano Pietronero and Sergi Valverde for their priceless suggestions.

\section*{Author contributions statement}

GA and MAJ designed and performed the research, as well as wrote the paper. 

\section*{Additional information}

\textbf{Competing financial interests} The authors declare no competing financial interests.

\end{document}